\documentclass[aps,prl,twocolumn,showpacs]{revtex4}
\usepackage{amssymb}
\usepackage{graphicx}

\newcommand{\be}{\begin{equation}}
\newcommand{\ee}{\end{equation}}
\newcommand{\ba}{\begin{eqnarray}}
\newcommand{\ea}{\end{eqnarray}}

\begin{document}

\title{\textbf{ Local and Effective Temperatures of Quantum Driven Systems}}
\author{Alvaro Caso, Liliana Arrachea and Gustavo S. Lozano}

\address{Departamento de F\'{\i}sica, FCEyN, Universidad de Buenos Aires,
Pabell\'on 1, Ciudad Universitaria, 1428, Buenos Aires, Argentina.}

\begin{abstract}
We introduce thermometers to define the local temperature of an electronic system driven out-of-equilibrium
by local AC fields. We also define the effective temperature in terms of a local fluctuation-dissipation-relation.
We show that within the weak driving regime these two temperatures coincide. We also discuss the behavior
of the local temperature along the sample. We show that it exhibits spatial fluctuations following an oscillatory
pattern. For weak driving, regions of the sample become heated, while others become cooled as a consequence of the
driving.
\end{abstract}

\maketitle

The study of heat transport at the meso and nano scale is being the subject of intense interest at present.
Motivation in this field is two-fold. On one side, the technological trend towards miniaturization of electronic
 circuits pushes for a better understanding of heat dissipation at this scale. From a more general point of view,
one is often faced with situations in which the very fundamental concepts of standard Statistical Mechanics
and Thermodynamics are put into test. This is for instance the case when the system under consideration is driven out
of equilibrium.

Dynamical evolution out of thermodynamical equilibrium takes place
in a great variety of  physical situations and many efforts have
been devoted during the last decade towards the
extension of standard thermodynamical concepts to this domain. Well
known examples in this area include  the ageing regime of glassy
systems, sheared glasses, granular materials and colloids. A break through in this field has been the
identification of {\em effective temperatures}, that is, even when
the system evolves out of the equilibrium, it is possible to
identify a parameter that has the same properties of the temperature
of a system at equilibrium. Even more, it is sometimes possible to
formulate a generalization of the equilibrium
 Fluctuation Dissipation Relations (FDR), with this new parameter playing the role of an effective temperature
\cite{cukupel,letoandco,letogus,lilileto}.

In the context of quantum transport, electronic devices driven under
AC potentials offer an ideal playground to explore these fundamental
issues. The study of heat transport in these systems has captured
increasing attention during the last years \cite{heatacdr,liliheat}.
Particularly appealing in this sense are setups where the AC fields
act locally within some region of the sample, that we define as the
``central system''. In practical configurations, this central (out of equilibrium) system
 is in contact with macroscopic wires which
remain at thermodynamical equilibrium and act as particle
and thermal reservoirs. A paradigmatic example is a quantum dot
driven at its walls by two voltages oscillating with a phase-lag
named ``quantum pump'' \cite{switkes,adia,liliflo}. Another example corresponds
to arrays of driven quantum capacitors coupled to the edge state of
an electronic gas in the Hall regime \cite{qcap1,qcap2}.

The aim of this work is to introduce the concept of {\em
local temperature} along the central system. To this end we follow
a procedure inspired in
 a pioneer work by Engquist and Anderson \cite{engq-an}. The idea is to include in the microscopic description
of the driven system a {\em thermometer}, namely,
 a macroscopic system which is  in local
thermodynamic equilibrium with the sample.
This theoretical construction enable us the investigation of interesting features on the behavior of the energy propagation
along the sample.
One of the most remarkable features is the development of
 spatial fluctuations of the local temperature, which leads to  local cooling of regions of the sample within the weak driving regime.
We also make a step further by identifying a FDR, which for weak driving casts an effective temperature
that is shown to exactly coincide with the local one \begin{small} measured \end{small} by the thermometer.

Our setup, including the device with the driven system in contact to reservoirs and thermometer is
described by the  Hamiltonian:
\ba
H(t)&=& H_{sys}(t)+H_{cP}+H_{P},\nonumber\\
H_{sys}(t)&=& H_{L}+H_{cL}+ H_{C}(t)+H_{cR}+ H_{R}. \ea The piece
$H_{sys}(t)$ contains the term describing the  central system ($C$)
with the AC fields, $H_C(t)=H_0+H_V(t)$ as well as terms
corresponding to  left ($L$)  and right ($R$) reservoirs with the
ensuing contacts $H_{cL}$ and $H_{cR}$. The term $H_P$ represents
the thermometer. It consists in a macroscopic system 
weakly coupled to a given point $l_P$ of $C$,
through a contact described by $H_{cP}$. It behaves like a reservoir
with a temperature $T_{lP}$ that is determined by the condition of a vanishing
heat flow between it and $C$. This is the
thermal counterpart of a voltage probe (see \cite{engq-an,friedel,
voltprobe}). All the reservoirs are modeled by systems of
non-interacting electrons with many degrees of freedom: $H_{\alpha}=
\sum_{k \alpha } \varepsilon_{k \alpha} c^{\dagger}_{k \alpha} c_{k
\alpha}$, being $\alpha=L,R,P$. The corresponding contacts are $H_{c
\alpha}= w_{c \alpha} (c^{\dagger}_{k \alpha} c_{l
\alpha}+c^{\dagger}_{l \alpha} c_{k \alpha})$, where $l \alpha$
denotes the coordinate of $C$ at which the reservoir $\alpha$ is
connected. We take into account the non-invasive property of the
thermometer \cite{engq-an}
 by treating $w_{cP}$ at the lowest order of perturbation theory when necessary. We leave for the moment $H_C$ undetermined 
as much of the
coming discussion is model independent.

The dynamics of the system is best described within the Schwinger-Keldysh Green functions formalism. This involves
 the calculation of the Keldysh and retarded Green functions,
\begin{eqnarray}
G^{K}_{l,l'}(t,t')&=& i \langle c^{\dagger}_{l'}(t') c_l(t) - c_l(t) c^{\dagger}_{l'}(t')  \rangle,
\nonumber \\
G^R_{l,l'}(t,t')&=& -i \Theta(t-t') \langle c_l(t) c^{\dagger}_{l'}(t')+c^{\dagger}_{l'}(t') c_l(t)
\rangle ,
\label{green}
\end{eqnarray}
where the indexes $l,l'$ denote spatial coordinates of the central system.
These Green functions can be evaluated after solving the Dyson equations.
For AC driven systems, it is convenient to use the
Floquet Fourier representation of these functions \cite{liliflo}:
\begin{equation}
 G_{l,l'}^{K,R}(t, t-\tau) = \sum_{k=-\infty}^{\infty}  \int_{-\infty}^{\infty} \frac{d\omega}{2 \pi}
 e^{-i (k \Omega_0 t + \omega \tau)}  G_{l,l'}^{K,R}(k,\omega),
\end{equation}
where $\Omega_0$ is the frequency of the AC fields.

We determine the local temperature by requiring that the heat current from the system to the thermometer vanishes. We will assume
here that the L and R leads are at the same temperature $T$ and that both leads and the thermometer have the same chemical
potential $\mu$. We work in units where $\hbar=e=k_{B}=1$.
As shown in \cite{liliheat}, given $H_C(t)$ without many-body interactions,
the heat current from the central system to the thermometer can be
expressed as
\ba
& &J_P^Q = \sum_{\alpha=L,R,P} \sum_{k=-\infty}^{\infty} \int_{-\infty}^{\infty} \frac{d\omega}{2 \pi}
 \{  [f_\alpha(\omega)-f_P(\omega_k)] \nonumber \\
& &   \times (\omega_k - \mu)
\Gamma_P(\omega_k) \Gamma_\alpha(\omega) \left| G^R_{lP,l\alpha}(k,\omega)\right|^2
  \} \label{jq}
\ea
where $\omega_k=\omega+k\Omega_0$ and
$\Gamma_{\alpha}(\omega) = -2 \pi |w_{\alpha}|^2 \sum_{k \alpha} \delta(\omega-\varepsilon_{k \alpha})$ are
the spectral functions that determine the escape to the reservoirs, and $f_\alpha(\omega)=
1/[e^{\beta_{\alpha}(\omega -\mu)}+1]$, is the Fermi function, which depends on the temperature
$T_{\alpha}=1/\beta_{\alpha}$ and the chemical potential  of the reservoir $\alpha$.
 Thus, the local temperature
$T_{lP}$ corresponds to the solution of the equation $J_P^Q(T_{lP}) =
0$. In general, the exact solution must  be found numerically,
however, an exact analytical expression can be obtained within the
weak-coupling and low-temperature $T$ regime.

Before doing so, let us   analyze a FDR between the local Green
functions $ G_{l,l}^{K,R}(t, t')$. Let us recall that for systems in
equilibrium, the Fluctuation Dissipation Theorem establishes a
relation between the Keldysh (correlation) and Retarded Green
functions. Indeed, for a system like the one under consideration,
but without the time-dependent fields, it can be shown that
 the relation between the fluctuations in the system,  $iG_{l,l}^{0,K}(\omega)$,
 with the dissipation term
of the bath, $\Gamma_{\alpha}(\omega)$, is \cite{letogus,lilileto}:
\ba \!\!\!\!\! & &iG_{l,l}^{0,K}(\omega)=
\tanh[\frac{\beta(\omega-\mu)}{2}] \varphi^0_l(\omega), \label{fdt1}
\\
\!\!\!\!\!\!
& & \varphi^0_l(\omega)=-2
\mbox{Im}[G_{l,l}^{0,R}(\omega)]= \sum_{\alpha=L,R} |G^{0,R}_{l,l
\alpha}(\omega)|^2 \Gamma_{\alpha}(\omega)
 \ea
 where the supraindex zero indicates that
we are considering 
$H_V(t)=0$ and all the reservoirs at the same temperature
$T$. When the time-dependent term is turned on, identities
between Green functions \cite{liliflo} generalize to
 \ba iG^K_{l,l}(0,\omega)&=&
\sum_{k = -\infty}^{\infty} \tanh [\frac{\beta(\omega_{-k} -
\mu)}{2} ] \varphi_l(k,\omega_{-k})\label{fdtg2} \\
 \varphi_{l}(k,
\omega) & = &  \sum_{\alpha= L,R}
 \left|G^R_{l, l \alpha}(k,\omega)\right|^2 \Gamma_{\alpha}(\omega).
\ea 
We will show below that within the weak driving-adiabatic regime, where
 the term $H_V(t)$ is treated as a perturbation and driving frequency is smaller than the dwell
time of the electrons within the central system \cite{adia}, it
is possible to define an effective temperature
$T^{eff}_l=1/\beta^{eff}_l$ through the following relation: \ba
iG^K_{l,l}(0,\omega)-iG^K_{l,l}(0,\mu)& = &
\tanh[\frac{\beta^{eff}_l (\omega-\mu)}{2}] \overline{\varphi_{l}}(
\omega), \label{tef} \ea with $\overline{\varphi_{l}}( \omega)=-2
\mbox{Im} [G_{l,l}^R(0,\omega)]= \sum_k \varphi_l(k,\omega_{-k})$.
 A similar relation in the time-domain has
been studied numerically for a driven ring in contact to a reservoir
\cite{lilileto}. In the present problem, we are able to demonstrate
that for weak driving and low temperature, $T^{eff}_l$ coincides
with the temperature $T_{lP}$ determined by the thermometer.

We now turn to analyze in detail the weak driving  regime where we
consider $H_V(t)$ as a perturbation in evaluating $J_P^Q$ (see
\cite{liliflo,liliheat,voltprobe}). For reservoirs at low  temperature $T$
(compared with the Fermi energy), a Sommerfeld expansion may be
applied in  Eq. (\ref{jq}) leading to  \be
T_{lP}^2 \sim \frac{6}{\pi^2} \frac{ \sum_k \Phi_{lP}(k)} { \sum_k
F_{lP}(k,\mu_{-k})} + T^2 \frac{ \sum_k F_{lP}(k,\mu) }{ \sum_k
F_{lP} ( k,\mu_{-k} ) }, \label{tp1} \label{tp} \ee 
where
 \ba
\Phi_{l}(k) &&= \int_{\mu_{-k}}^{\mu} d\omega (\omega_k - \mu)\varphi_{l}(k,\omega), \\
 F_{l}(k,\omega)& &=\frac{d}{d\omega} [(\omega_k - \mu)\varphi_{l}(k,\omega)],
\ea
encode the dependence on the driving field and the geometry 
of the central system.
This expression makes it explicit the fact that the local temperature is different from the
temperature of the leads.

In order to be more specific, let us consider a driving term of the form:
\be \label{hv}
 H_V(t)= \sum_{j=1}^M V_j(t) c^\dagger_{lj} c_{lj} ,
\ee
 with
$V_j(t)= V_0 \cos( \Omega_0 t + \delta_j)$, being $lj$ the positions at where the AC fields are applied.
For small $V_0$ the Dyson equation is solved to lowest order in this amplitude and the only nonvanishing  Floquet
components of $G^R_{l,l'}(k,\omega)$ are those with $k=-1,0,1$
\cite{liliflo,liliheat}. 
 The adiabatic condition  is introduced
by expanding all  terms of (\ref{tp1}) in powers of $\Omega_0$. Keeping terms up to $\Omega_0^2$,
\ba
T_{lP}^2 &  & \sim T^2+ \frac{3}{\pi^2} \lambda_{lP}^{(0)}(\mu) \Omega_0^2 +  2
\lambda_{lP}^{(1)}(\mu) T^2 \Omega_0
\nonumber \\
& & - \frac{1}{2} \lambda_{lP}^{(2)}(\mu) T^2 \Omega_0^2,
 \label{tloc}
\ea
being
\be
\lambda_{l}^{(n)}(\omega)=  \frac{1}{ \sum_{k=-1}^{1} \varphi_{l}(k,\omega)}
\sum_{k=-1}^{1} (k)^{n+2} \frac{d^n[ \varphi_{l}(k,\omega)]}{d \omega^n} .
\ee

We now carry out a similar analysis with the effective temperature.
For weak driving, the relevant  electronic energies $\omega$ are
such that $|\omega-\mu| \lesssim \mbox{max}(T,\Omega_0)$. For small $T$ and $\Omega_0$ we, thus, expand
both sides of Eq. (\ref{tef}) around $\omega=\mu$. Keeping terms up
to first order in $\omega$, we find:
\begin{equation}
 T^{eff}_l=\frac{1}{2}\frac{ \overline{\varphi}_{l}(\mu)}{\frac{d}{d\omega}
 \left[ \sum_{k=-1}^{1} \varphi_{l}(k,\omega_{-k})\tanh [\frac{\beta(\omega_{-k} - \mu)}{2}] \right]_{\omega=\mu} },
\end{equation}
which, for low driving frequency $\Omega_0$, reduces to:
\begin{equation} \label{som}
 T^{eff}_l=T [1+ \lambda_{l}^{(1)}(\mu) \Omega_0].
\end{equation}
By comparing this expression with (\ref{tloc}) we find that
$T^{eff}_{lP}=T_{lP}$ when $\Omega_0 \ll T$. We would like to
emphasize that our definition of local temperature is independent of
the weak driving-adiabatic assumption. However,  it is within this regime
where the system is \begin{small} slightly out of equilibrium \end{small} and the equivalence with the
effective temperature defined from the FDR (\ref{tef}) is expected.

In order to show other explicit results we will choose a simple set
up composed of  a central system with two local AC fields
oscillating with a phase-lag, i.e. $M=2$, $\delta_1=0$,
$\delta_2=\delta$ in Eq. (\ref{hv}), a simple
 model for a quantum pump. As  central system
we take a one-dimensional lattice of $N$ sites, with the
first one connected to $L$  and the site $N$ connected to $R$
($l\alpha=1,N$, for $\alpha=L,R$), and $H_0= - w \sum_{l=1}^{N-1}
(c^{\dagger}_l c_{l+1} + H.c.)$. All  energies and temperatures
 are expressed in units of $w$.

\begin{figure}
\centering
\includegraphics
[width=80mm,angle=0]{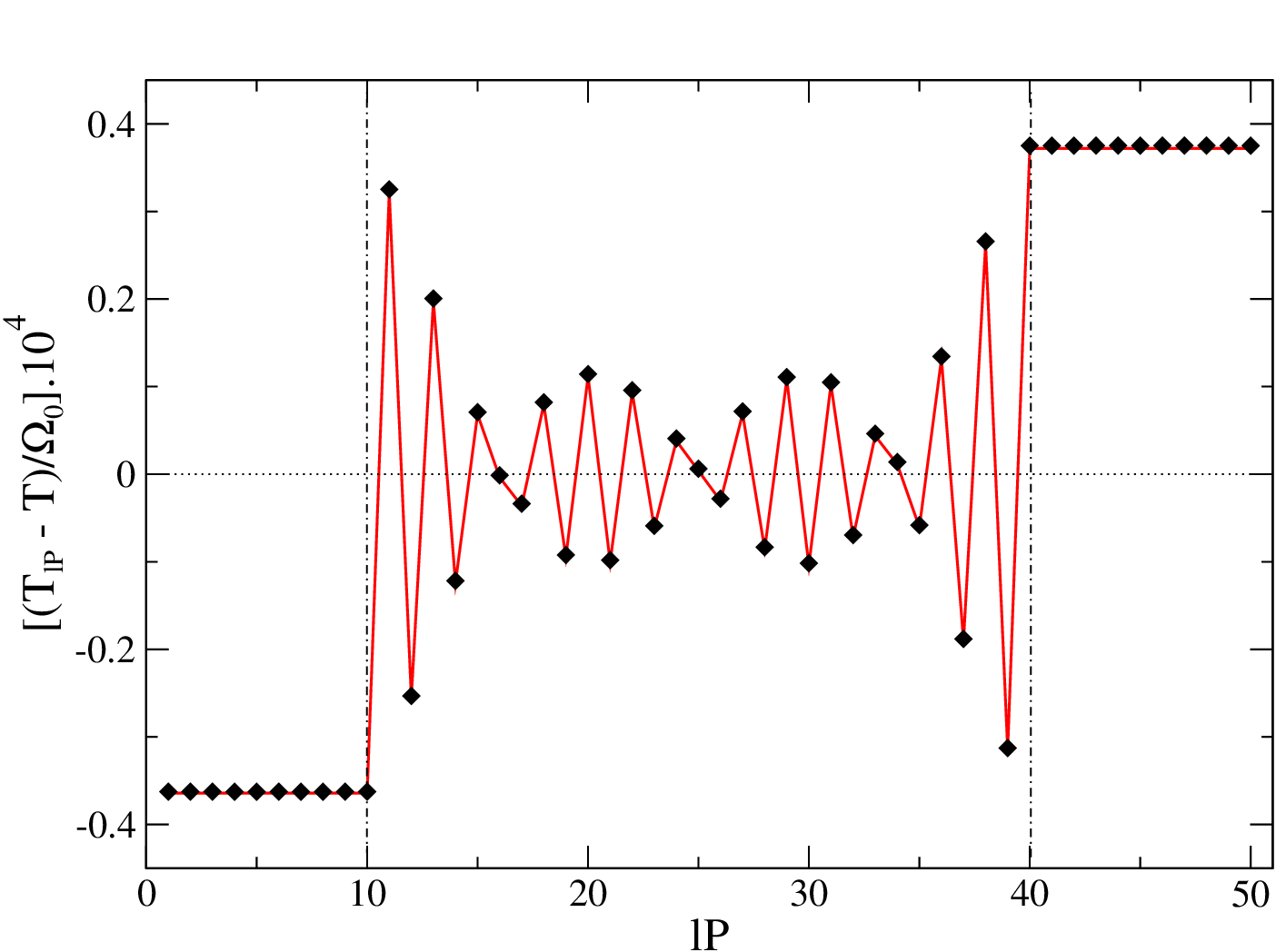} {\small {} }
\caption{{
(Color online) Local (solid red) and effective (black diamonds) temperatures along
a one-dimensional model of $N=50$ sites with two AC fields
operating with a phase-lag $\delta=\pi/2$ at the positions indicated in dotted lines.
The system is
in contact to reservoirs with chemical potentials
$\mu=0.2$, and temperature
$T=0.001$. The driving frequency is $\Omega_0=10^{-6}$ and the the amplitude is
$V_0 =0.05$. }}
\label{fig3}
\end{figure}

We  first discuss the behavior of the local temperature along the
system. On general grounds, one expects that the driving fields heat
the  sample giving rise to dissipation of energy from the system to
the reservoirs. Therefore,  intuition would indicate that the local
temperature of the sample is higher than that of the reservoirs. At
weak driving, however, it has been shown that it is possible to
coherently transport energy along the sample in a way that some of
the AC fields develop power against other ones with a low
dissipation of heat to the reservoirs \cite{liliheat}. Actually, the
mechanism of energy exchange between the fields behaves $\propto
\Omega_0$, while  the rate at which energy is dissipated as heat is
$\propto \Omega_0^2$. A non trivial behavior of the local
temperature, could, thus, take place within this regime. This is,
in fact, the case shown in Fig.~\ref{fig3}, where parameters
are chosen within the weak driving
and low temperature regime. In agreement
with our previous discussion we show that $T_{lP} \equiv T^{eff}_{lP}$.  Remarkably, as a
function of the position at which the thermometer is connected or,
equivalently, the FDR (\ref{tef}) is evaluated, the local
temperature varies along the sample, being lower than $T$ in some
places, while higher in others. Between the two pumping centers, the
local temperature displays oscillations with a spatial period $\sim
2 k_F$. These oscillations are due to processes $\propto V_0^2
\Omega_0 \sin (\delta)$ and have a similar origin as the Friedel
oscillations detected by voltage probes \cite{friedel}. In the
present system, the latter are a consequence of interference
processes in the charge transport  that take place between the two
pumping centers, which act as dynamical local impurities
\cite{voltprobe}. Therefore, the oscillatory behavior in $T_{lP}$ is
a signature of the coherence of at least some component of energy
transport along the sample.

\begin{figure}
\centering
\includegraphics
[width=80mm,angle=0]{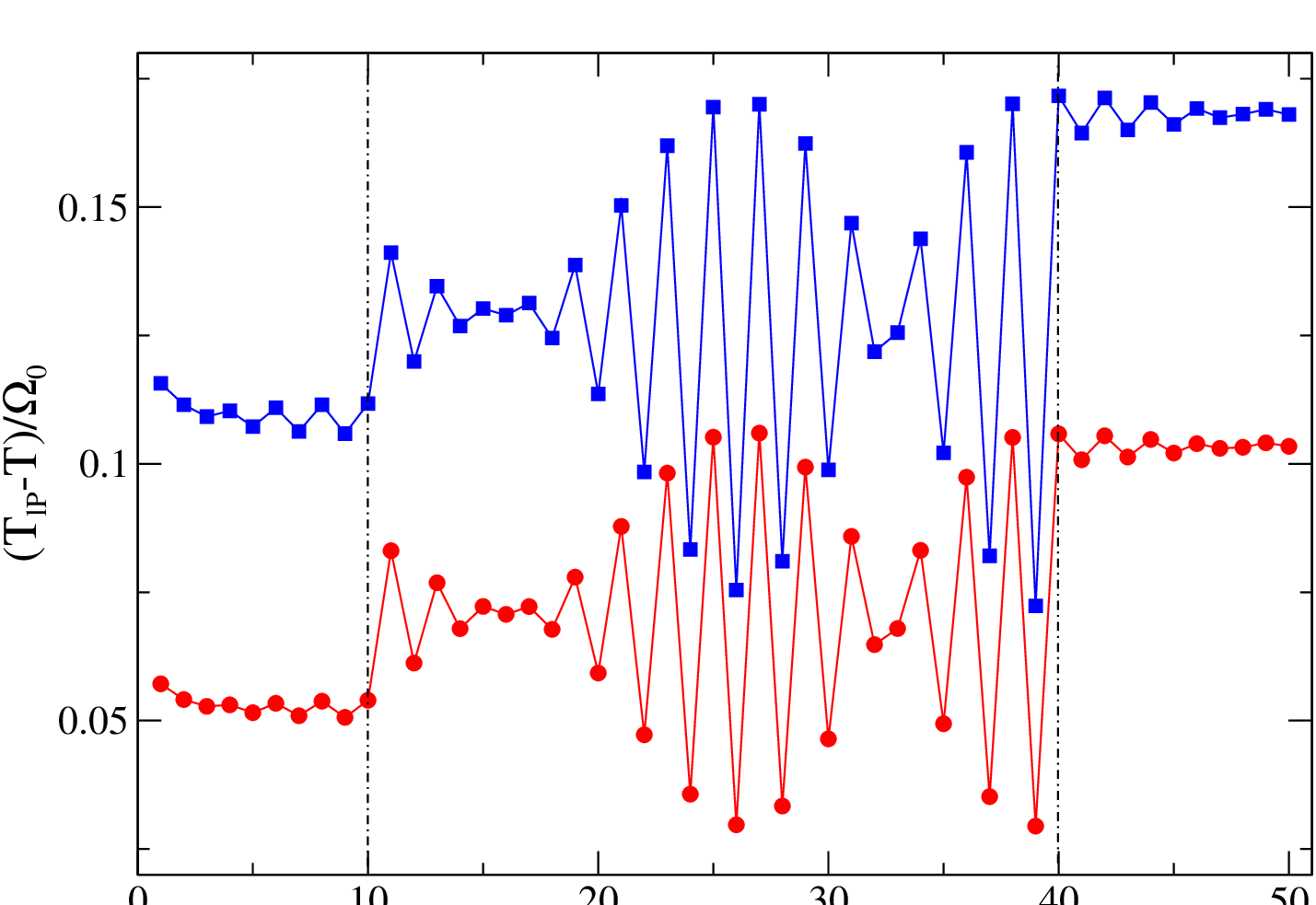} {\small {} }
\caption{{
(Color online) Local temperature along the sample for $\mu=0.2$,
$\delta$=$\pi/4$, $V_0=0.25$, $\Omega_0=0.1$ and
$T=0.001$ (blue squares) and $T=0.01$ (red circles)}.
All other parameters are the same as in Fig. \ref{fig3}.}
\label{fig1}
\end{figure}

In Fig. \ref{fig1} we show results for the local temperature as a function
of the position of the thermometer beyond
the regime of validity of the weak driving and/or adiabatic approximation.
 In all  cases, oscillations of the local temperature between the two pumping centers are apparent,
 pointing to
the survival of a coherent component in the energy propagation. At fixed
$T$, the mean temperature of the sample,
 $T_m =
(\sum_{lP=1}^N T_{lP})/N$ grows as  $\Omega_0$ increases
and  becomes soon higher than the temperature of the reservoir.

\begin{figure}
\centering
\includegraphics
[width=80mm,angle=0]{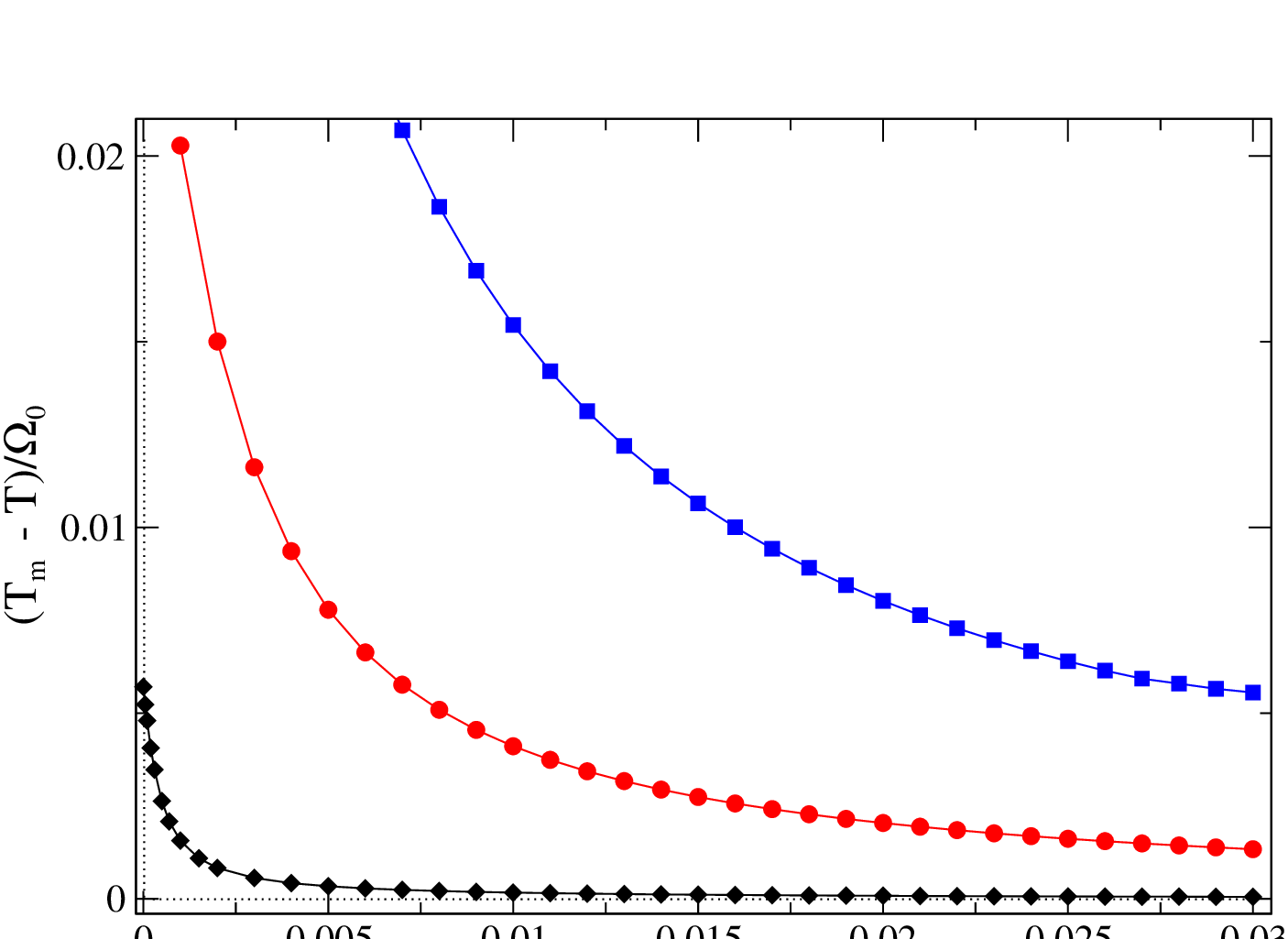} {\small {} }
\caption{{
(Color online) Departure of the mean temperature of the central system
from the temperature of the reservoirs for
$\mu=0.2$,
$\Omega_0=0.1$,
$\delta=\pi/4$, $V_0=0.1$ (blue squares), $V_0 =0.05$ (red circles),
$V_0 =0.01$ (black diamonds). Other parameters are the same as in Fig. \ref{fig3}.}}
\label{fig2}
\end{figure}

Finally, we analyze the behavior of the mean temperature $T_m$ as a function
of $T$. Results are shown in Fig. \ref{fig2} for a driving frequency beyond the
adiabatic regime and different values of the pumping amplitude $V_0$. As
expected, for  $T$ fixed, the mean temperature of the sample
 increases for increasing $V_0$. Instead, $T_m-T$ is a decreasing function of $T$.
This reflects the fact that for reservoirs at a high temperature, the
effect of the driving becomes washed up and the sample becomes mainly heated due to the
contact with a high temperature environment.

To conclude, we have defined local temperature for a quantum system driven out of equilibrium. The behavior of this quantity indicates
a global heating, which manifests itself in the form of a mean temperature $T_m$
higher than the one of the reservoirs. A more striking feature is the occurrence of  $2k_F$ oscillations
in the local temperature. This is an indication of quantum interference, i.e. coherence in the energy
propagation along the sample. At weak driving, these oscillations give place to the local cooling of the sample.
 We have also defined an effective temperature
from a local
fluctuation dissipation relation. We have shown that for weak driving and
for temperatures smaller than the Fermi energy of electrons, the latter coincides
with the one defined by the the thermometer. This equivalence has been previously
established only
for classical spin systems \cite{cukupel}. The fact that  
such a kind of equivalence holds for quantum fermionic systems is an
important conceptual issue and its scope for other systems worth further future
investigation.

We acknowledge support from CONICET and UBACYT, Argentina.


\end{document}